**Observation of Local Pairs through Nonequilibrium Terahertz Spectroscopy in $YBa_2Cu_3O_{7-\delta}$ Below $T_c$**

Jinzhong Zhang[1], Qingming Huang[1], Zhangqiang Yang[1], Zhiyuan Sun[2], Ye Yang[1]

1. State Key Laboratory of Physical Chemistry of Solid Surfaces, College of Chemistry and Chemical Engineering, Xiamen University, Xiamen 361005, China
2. Department of Physics, Tsinghua University, Beijing 100084, China

**Abstract:** Pairing states are essential to understanding the underlying mechanisms of high-temperature superconductivity. Here the non-superconducting state in an optimally doped $\mathrm{YBa_2Cu_3O_{7-\delta}}$ film was driven out of equilibrium by an optical pump with low fluence at a temperature well below the critical temperature ($T_c$), and its recovery dynamics were exclusively measured using transient terahertz spectroscopy. The pump-fluence dependent experiments unveiled evidence of local pairs without superconductivity coexisting with the superconducting Cooper pairs. An energy gap opening induced by the local pairing with short-range coherence was invoked to rationalize the temperature-dependent recovery time of local pairs with a characteristic divergence at a temperature substantially below $T_c$. These local pairs displayed remarkable likeness to the short-range pair-density-wave state. Our finding sheds light on understanding the dynamic interplay between the coexisting superconducting and non-superconducting pairs in cuprate superconductors.

Cuprate superconductors exhibit critical temperatures significantly higher than conventional superconductors, which has spurred extensive research into understanding their underlying mechanisms. One of the central questions revolves around the pairing mechanism responsible for the formation of Cooper pairs, which is fundamentally different from the pairing in conventional superconductors described by Bardeen-Cooper-Schrieffer (BCS) theory.(*1*) The Cooper pairs can condense into a superfluid with long-range coherence and thus give rise to superconductivity. In conventional BCS superconductors, the formation of Cooper pairs and the establishment of long-range coherence occur simultaneously below the critical temperature ($T_c$). However, these two quantum phenomena take place separately in the cuprate (particularly in the underdoped ones) and other unconventional SCs.(*2-7*) In these superconductors, pairing may also lead to Cooper pairs without superconductivity, i.e., without long-range phase coherence. It has been demonstrated that electron pairs preform at temperatures above $T_c$ via local pairing, which might account for the opening of a pseudogap.(*2, 8*) At a temperature well below $T_c$, signatures of local pairs without long-range coherence have also been observed using scanning tunneling microscopy,(*9-12*) and the charge order revealed by this technique is conjectured to be intimately related to a short-range pair-density-wave (PDW) state coexisting with the d-wave superconductivity.(*13-18*) However, it is very challenging to measure the interplay between superconducting pairs and local pairs below $T_c$ using the steady state experimental tools because a fine balance is established between them in thermal equilibrium. Inspired by the time-resolved spectroscopic studies on superconductors, the recovery dynamics from a light-induced nonequilibrium state can provide insights into the interplay between the coexisting states. (*19-41*)

Since the superconducting and non-superconducting states exhibit distinct frequency-dependent conductivities, the transient terahertz (THz) spectroscopy can be exploited to separately track their respective restoration dynamics after optical excitation. Here, using this technique, we uncovered the pairwise recombination of nonequilibrium quasiparticles in an optimally doped $\mathrm{YBa_2Cu_3O_{7-\delta}}$ (YBCO) film. At temperature well below $T_c$, the transient THz response for low pump fluence primarily arises from the partial depletion of the non-superconducting state whose recovery proceeds in a pairwise fashion, indicative of the existence of local pairs at the ground state. The temperature dependence of the time constant for the non-superconducting state recovery shows a clear divergence near 70 K (notably below the $T_c$), implying an energy gap opening associated with local pairing. These local pairs display remarkable likeness to the short-range PDW state coexisting with the superconducting state.

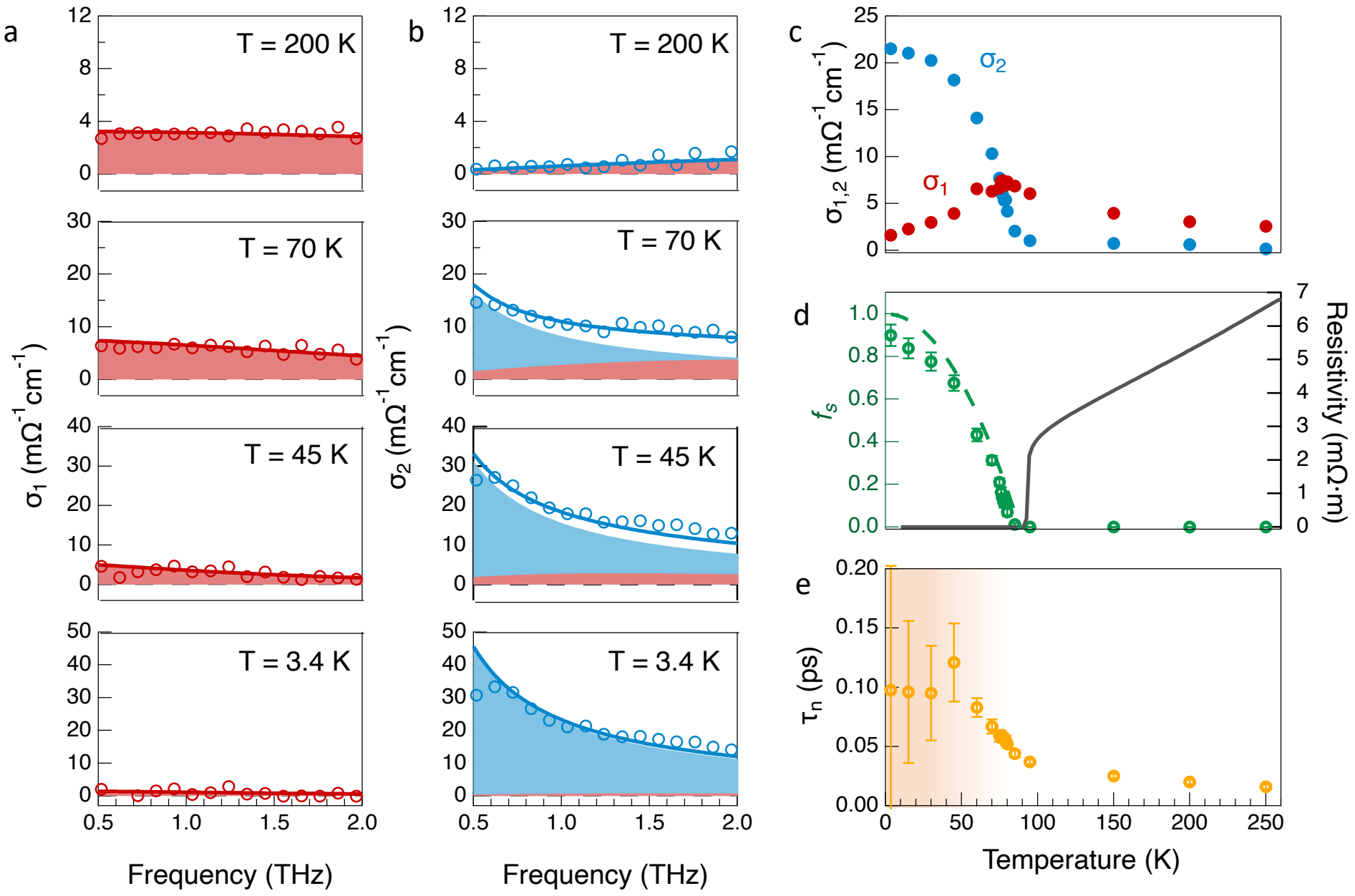

**Figure 1. Temperature dependence of terahertz conductivities at equilibrium condition.** (a) Real ($\sigma_1$) and (b) imaginary ($\sigma_2$) parts of the YBCO equilibrium conductivity of measured at representative temperatures (open circles). Solid curves represent the two-fluid model. The red and blue shades represent the Drude and superfluid components, respectively. (c) Temperature dependence of $\sigma_1$ and $\sigma_2$ magnitudes averaged from 0.9 to 1.2 THz. (d) The measured (green circles) and calculated (green dash curve) superfluid fraction, $f_s$, at various temperatures. The black solid curve shows the electric resistivity as function of temperature. (e) Temperature dependent scattering time of the normal fluid.

The $YBa_2Cu_3O_{7-\delta}$ ($\delta \approx 0.1$) thin film was deposited on a MgO substrate using magnetron sputtering, and the critical temperature for this sample is determined to be 91 K (Fig. S1). The complex conductivities ($\tilde{\sigma}$) of the YBCO film over the frequency range of 0.5 to 2.0 THz were measured using the equilibrium THz spectroscopy at various temperatures (Fig. S2). At a temperature well above $T_c$ (e.g., $T$ = 200 K), the real and imaginary parts of $\tilde{\sigma}$, denoted as $\sigma_1$ and $\sigma_2$, respectively, show typical Drude responses, characteristic of metallic conductivity (Fig. 1a and 1b). As the temperature drops below $T_c$, $\sigma_1$ is fading away because the spectral weight tends to condense into the delta function at zero frequency, while $\sigma_2$ shows a spectral weight transfer from the high frequency to the low frequency. For $T \ll T_c$ (e.g., $T$ = 3.4 K), $\sigma_2$ displays a typical $1/\omega$ dependence, a hallmark of superconductivity. The temperature dependence of $\tilde{\sigma}$ for YBCO observed here aligns well with the literature,(*42-46*) which can be described by the phenomenological two-fluid model:

$$\tilde{\sigma}(T) = \rho_n(T)\frac{\tau_n(T)}{1 - i\omega\tau_n(T)} + \rho_s(T)\frac{i}{\omega} \qquad (1)$$

where $\rho_n(T)$ and $\rho_s(T)$ are the effective densities of the normal fluid and superfluid, respectively, and $\tau_n$ is the scattering time for the normal fluid. The sum of $\rho_n(T)$ and $\rho_s(T)$ is constrained by the total number of electrons and assumed to be constant, and the fraction of superfluid, $f_s(T)$, is then defined as $\rho_s(T)$ / $[\rho_s(T) + \rho_n(T)]$. By substituting this expression into Eq. (1), the temperature dependence of $\tilde{\sigma}$ is found to be governed by the temperature-dependent $f_s(T)$ and $\tau_n(T)$, which are then set as free fitting parameters. The fitting curves (solid curves, Fig. 1a and 1b) excellently match the experimental data (symbols, Fig. 1a and 1b). This model indicates that below $T_c$, $\sigma_1$ comprises the pure Drude component, while $\sigma_2$ consists of both superfluid (blue shades, Fig. 1b) and Drude (red shades, Fig. 1b) contributions. The two-fluid model was also successfully applied to $\tilde{\sigma}$ at other temperatures (Fig. S2).

The temperature dependent magnitudes of $\sigma_1$ and $\sigma_2$ display dramatic changes near $T_c$ (Fig. 1c), consistent with the literature.(*43, 45-48*) The onset of the temperature dependent $f_s(T)$ extracted from the fitting corresponds to the beginning of condensation, coinciding with $T_c$ determined from temperature dependent electric resistivity (Fig. 1d). The temperature dependence of $f_s(T)$ basically obeys the relation $f_s(T) = 1 - (T/T_c)^2$ (green dash curve, Fig. 1e), which is typical for a dirty d-wave superconductor.(*45, 48, 49*) Nonetheless, the measured $f_s$ slightly deviates from the prediction as the temperature approaches zero. The temperature dependence of $\tau_n$ given by the fitting is displayed in Fig. 1e. Since $\sigma_1$ becomes negligibly small at temperatures close to zero, the limited signal-to-noise ratio renders the extracted scattering rates at these low temperatures unreliable, manifested as the large error bars. Despite the large fitting uncertainties, we can still qualitatively discern that $\tau_n$ rises steeply below $T_c$ and then saturate (orange shade, Fig. 1e). An abrupt increase of $\tau_n$ below $T_c$ was also observed by microwave spectroscopy in high-purity YBCO crystals and attributed to the suppression of the thermally excited quasiparticles.(*50*)

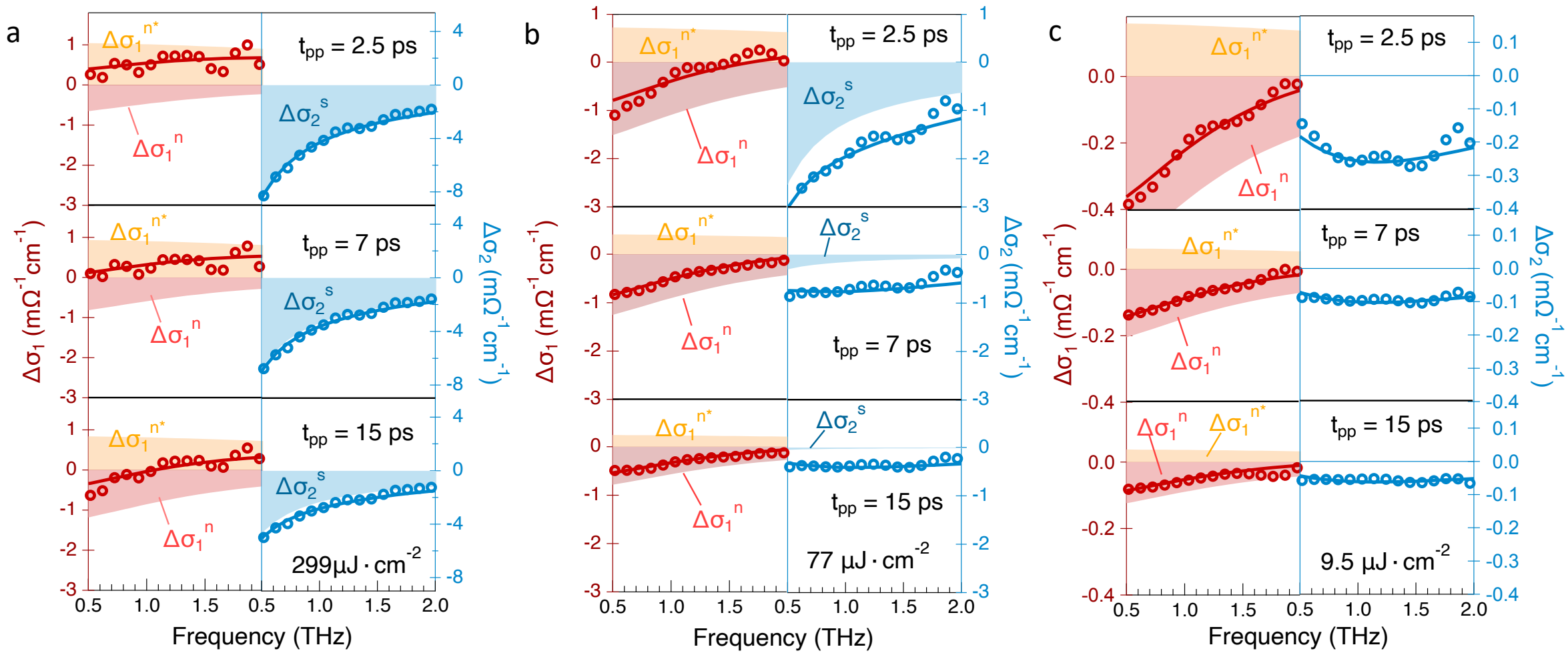

**Figure 2**. **Pump-fluence dependence of photoinduced THz conductivities ($\Delta\tilde{\sigma}$).** The real ($\Delta\sigma_1$) and imaginary ($\Delta\sigma_2$) parts of $\Delta\tilde{\sigma}$ for typical (a) high, (b) medium and (c) low pump fluences. The red and blue circles represent $\Delta\sigma_1$ and $\Delta\sigma_2$ recorded at indicated pump-THz delays ($t_{pp}$), respectively. The measurements were performed at 3.4 K to minimize the thermal excitation of Cooper pairs. The pump photon energy is 1.55 eV. Solid curves are the fits based on Eq 2. The orange, red and blue shades represent the $\Delta\sigma_1^{n^*}$, $\Delta\sigma_1^{n}$ and $\Delta\sigma_2^{s}$ (see main text for their definitions), respectively.

The dynamic behaviors of the carriers under nonequilibrium condition is investigated using transient THz spectroscopy. The experimental method is described in the Supplementary Material. Earlier transient optical reflection studies uncovered two quasiparticle relaxation components below $T_c$, ascribed to excitation of both the superfluid and normal fluid (i.e., pseudogap state).(*21, 39, 51, 52*) Thus, after taking into account the normal fluid excitation, the photoinduced change in $\tilde{\sigma}$ (denoted as $\Delta\tilde{\sigma}$) can be described by the following formula,

$$\Delta\tilde{\sigma} = -\Delta\rho_s\left(\frac{i}{\omega}\right) - \Delta\rho_n\left(\frac{\tau_n}{1 - i\omega\tau_n}\right) + (\Delta\rho_s + \Delta\rho_n)\left(\frac{\tau_n^*}{1 - i\omega\tau_n^*}\right) \qquad (2)$$

where $\Delta\rho_n$ and $\Delta\rho_s$ are the photoinduced changes in $\rho_n(T)$ and $\rho_s(T)$, respectively, and $\tau_n^*$ is the scattering time of the nonequilibrium quasiparticles created from the two-fluid excitation (denoted as QPs*). The equilibrium quasiparticles are then denoted as QPs$^0$. The first and second terms account for a conductivity reduction due to the depopulation of the superfluid and QPs$^0$, respectively, while the third term corresponds to a conductivity increase due to the photogenerated QPs*. As shown in Fig. 2, $\Delta\tilde{\sigma}$ at representative pump-THz delays ($t_{pp}$) exhibits strong dependence on the incident pump fluence ($I_p$). The extraction method of $\Delta\tilde{\sigma}$ is described in Supplementary Materials (Fig. S3). The measured $\Delta\tilde{\sigma}$ for different pump fluences can be simultaneously fitted by Eq. 2 (solid curves, Fig. 2) with linked fitting parameters of $\tau_n$ and $\tau_n^*$, revealing that the dependence of $\Delta\tilde{\sigma}$ on $I_p$ or $t_{pp}$ stems from the variation of $\Delta\rho_n$ and $\Delta\rho_s$. The values of $\tau_n$ and $\tau_n^*$ are determined to be 0.12$\pm$0.02 ps and 0.034$\pm$0.005 ps, respectively. We find that $\tau_n^*$ is much smaller than $\tau_n$, implying that the mobility of QPs* should be much lower than that of $QPs^0$, in line with the literature.(*53*) By substituting the fitting parameters into Eq. 2, the imaginary part of the first term ($\Delta\sigma_2^s$ ), the real part of the second term ($\Delta\sigma_1^n$ ) and the real part of the third term ($\Delta\sigma_1^{n^*}$) can be separately obtained (colored shades in Fig.2), which are exploited to trace the dynamics of the superfluid, $QP^0$ and QP* components, respectively.

The optical excitation with high $I_p$ leads to a negative $\Delta\sigma_2$ with a nearly $1/\omega$ dependence and a small positive $\Delta\sigma_1$ (Fig. 2a), ascribed to the partial depletion of superfluid and formation of QPs*, respectively.(*19, 20, 36, 45*) The fitting reveals that $\Delta\sigma_1$ comprises a contribution from $\Delta\sigma_1^n$ with a negative value due to a reduction in the $QP^0$ density after excitation. With increasing $t_{pp}$, $\Delta\sigma_2^s$ and $\Delta\sigma_1^{n^*}$ decay concurrently because QPs* condense into the superfluid (i.e., recovery of the superfluid), and this decay is accompanied by a growth of $\Delta\sigma_1^n$, indicating a continuous depletion of $QPs^0$. As $I_p$ drops to a lower level (Fig. 2b), $\Delta\sigma_2$ is initially dominated by $\Delta\sigma_2^s$ and then quickly evolves into a Drude response with increasing $t_{pp}$. The decay of $\Delta\sigma_2^s$ observed here is remarkably faster than that in the higher-$I_p$ case, indicative of a quicker recovery of the superfluid density. On the contrary, the $QP^0$ density recovers at a relatively slow pace, evidenced by the relatively slow decay of $\Delta\sigma_1^n$. When $I_p$ falls into a typical low-fluence regime (Fig. 2c), both $\Delta\sigma_1$ and $\Delta\sigma_2$ resemble the Drude response, and $\Delta\sigma_2^s$ can no longer be observed in $\Delta\sigma_2$ even at the short $t_{pp}$. The absence of $\Delta\sigma_2^s$ implies that the superfluid density remains intact after such a weak excitation, and the photoconductivity change solely arises from the excitation of $QPs^0$. With increasing $t_{pp}$, the decay of $\Delta\sigma_1$ or $\Delta\sigma_2$ simply reflects the relaxation of QPs* into $QPs^0$. While $\Delta\tilde{\sigma}(\omega, t_{pp})$ extracted from the THz spectroscopy at short delays could be distorted, especially for carriers with large momentum relaxation time,(*32, 54, 55*) the scattering time of the photogenerated QPs* is much shorter than the pump-probe delay, and thus the aforementioned distortion in $\Delta\tilde{\sigma}(\omega, t_{pp})$ should be negligible.

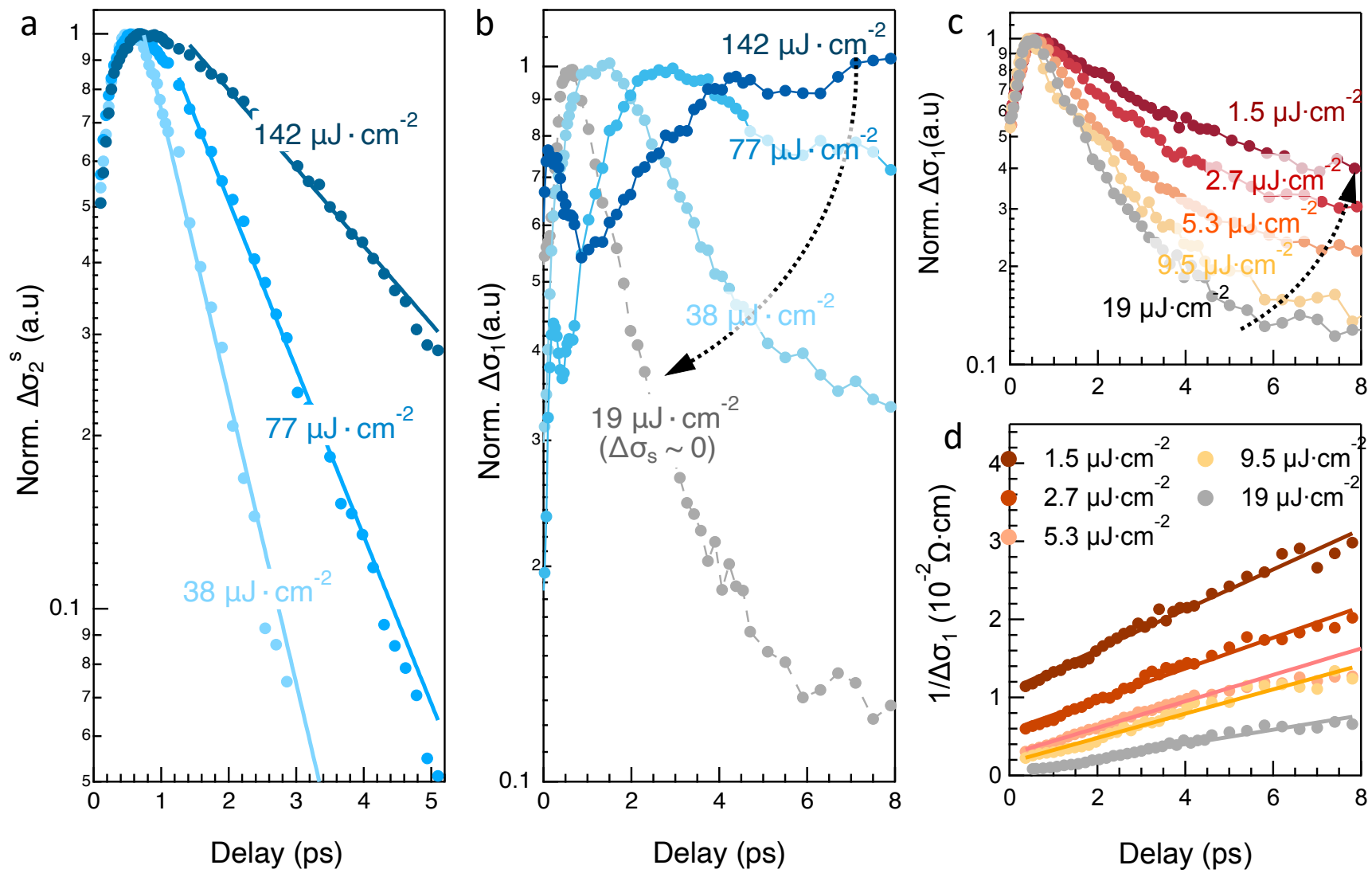

**Figure 3. Recovery dynamics of superconducting and non-superconducting states after excitation.** Normalized kinetics of (a) the superfluid component ($\Delta\sigma_2^s$) and (b) Drude component ($\Delta\sigma_1$) for typical high pump fluences. Note that the sign of the kinetics is changed after normalization. The lines in panel (a) are the monoexponential fits. The trace corresponding to 19 $\mu J \cdot cm^2$ in panel (b) is plotted to illustrate the $I_p$-dependent decay trend, and under this pump fluence $\Delta\sigma_2^s$ is negligible. The dash curve with an arrow head is a visual aid for decreasing $I_p$. (c) Normalized $\Delta\sigma_1$ kinetics for the indicated pump fluences in the low-$I_p$ regime. (d) The reciprocal of $\Delta\sigma_1$ ($1/\Delta\sigma_1$) plotted versus the delay time. The solid lines represent fits based on the biparticle annihilation model, and their slopes are equal to the rate constants. These kinetics were all measured at 3.4 K.

As $\Delta\sigma_2^s$ only emerges when $I_p$ exceeds a certain level at temperature well below $T_c$ (e.g., 3.4 K). Above this critical $I_p$, the kinetics of $\Delta\sigma_2^s$ for different $I_p$ are extracted through fitting $\Delta\tilde{\sigma}$ at different $t_{pp}$ to explore the $I_p$-dependent superfluid recovery dynamics (Fig. 3a). The $\Delta\sigma_2^s$ kinetics are well described by a single-exponential function with rate constant decreasing with increasing $I_p$, which can be interpreted by the Rothwarf-Taylor (R-T) model.(*19, 24, 56, 57*) Although the R-T model is originally derived for isotropic s-wave superconductors, it has also been widely exploited to deal with condensation dynamics of the nonequilibrium antinodal quasiparticles in d-wave superconductors.(*19, 24, 56, 57*) In this model, pairing and pair breaking are mediated by high-frequency phonons (HFPs) with energy greater than the binding energy of the pairs, 2Δ (Δ is the superconducting energy gap), and the apparent decay of $\Delta\sigma_2^s$ kinetics is then masked by the relatively slow anharmonic decay of the nonthermal HFPs arising from hot carrier relaxation because a quasi-equilibrium is established between the QPs* and HFPs under high-$I_p$ conditions.(*19, 24, 36, 56*) Unlike the kinetics of $\Delta\sigma_2^s$, the kinetics of $\Delta\sigma_1$ (Fig. 3a) consist of a relatively slow growth followed by a decay component, and both the growth and decay rates decrease with increasing $I_p$. The growth of $\Delta\sigma_1$ corresponds to an increase in the QP* density. The QPs* can either be formed by destroying the superfluid or exciting the coexisting QPs$^0$. The former mechanism is excluded because the growth of $\Delta\sigma_1$ is accompanied by the decay of $\Delta\sigma_2^s$ (i.e., the recovery of the superfluid that is supposed to drain the QPs* rather than increase its density). Therefore, the excitation of QPs$^0$ should account for the growth of $\Delta\sigma_1$ during the superconductivity recovery.

As demonstrated above, $\Delta\sigma_2^s$ is absent when $I_p$ falls into the aforementioned low-$I_p$ regime (Fig. 2c), suggesting that the superfluid density is intact under very weak excitation. In stark contrast with the high-$I_p$ scenario, the kinetics of $\Delta\sigma_1$ in the low-$I_p$ regime decays faster with increasing $I_p$ (Fig. 3c), and the reciprocal of $\Delta\sigma_1$ kinetics ($1/\Delta\sigma_1$) displays a linear relationship with $t_{pp}$ (Fig. 3d), a hallmark of the biparticle recombination pattern described by the following equation,

$$\frac{1}{\Delta\sigma_1(t_{pp})} = \frac{1}{\Delta\sigma_1(0)} + k_2 t_{pp} \qquad (3)$$

where the slope, $k_2$, is the rate constant for the biparticle decay. Although the recovery of the superfluid component could show biparticle recombination that was interpreted within the framework of the RT model, (*19*) the kinetic traces in Fig. 3c and Fig. 3d suggests a distinctly different recombination mechanism because they represent the normal fluid recovery. In such a scenario, their biparticle decay pattern evidences that the QPs* return to the non-superconducting ground state (i.e., QPs$^0$) through recombining into pairs. In other words, QPs$^0$ may exist as pairs without long-range phase coherence (i.e., local pairs) at 3.4 K. Thus, the very weak optical excitation breaks these local pairs rather than the superconducting pairs, and the $\Delta\sigma_1$ recovery represents the local pairing process. This pairwise recovering for the non-superconducting state are observed for optical pumps with wavelength ranging from ultraviolet to infrared (Fig. S4), suggesting that this phenomenon is independent of the pump photon energy and optical penetration depth.

Because the superconducting energy gap is closed above $T_c$, the $\Delta\sigma_1$ kinetics measured above $T_c$ reflect the intraband thermalization of QPs*, which include a fast component due to cooling via carrier-phonon interaction and a slow component due to heat dissipation (Fig. S5). The cooling rate is independent of $I_p$, while the heat effect becomes more significant with increasing $I_p$. Because of the distinctly different decay behaviors, the intraband relaxation of QPs* can be excluded as a probability to account for the $\Delta\sigma_1$ kinetics shown in Fig. 3d.

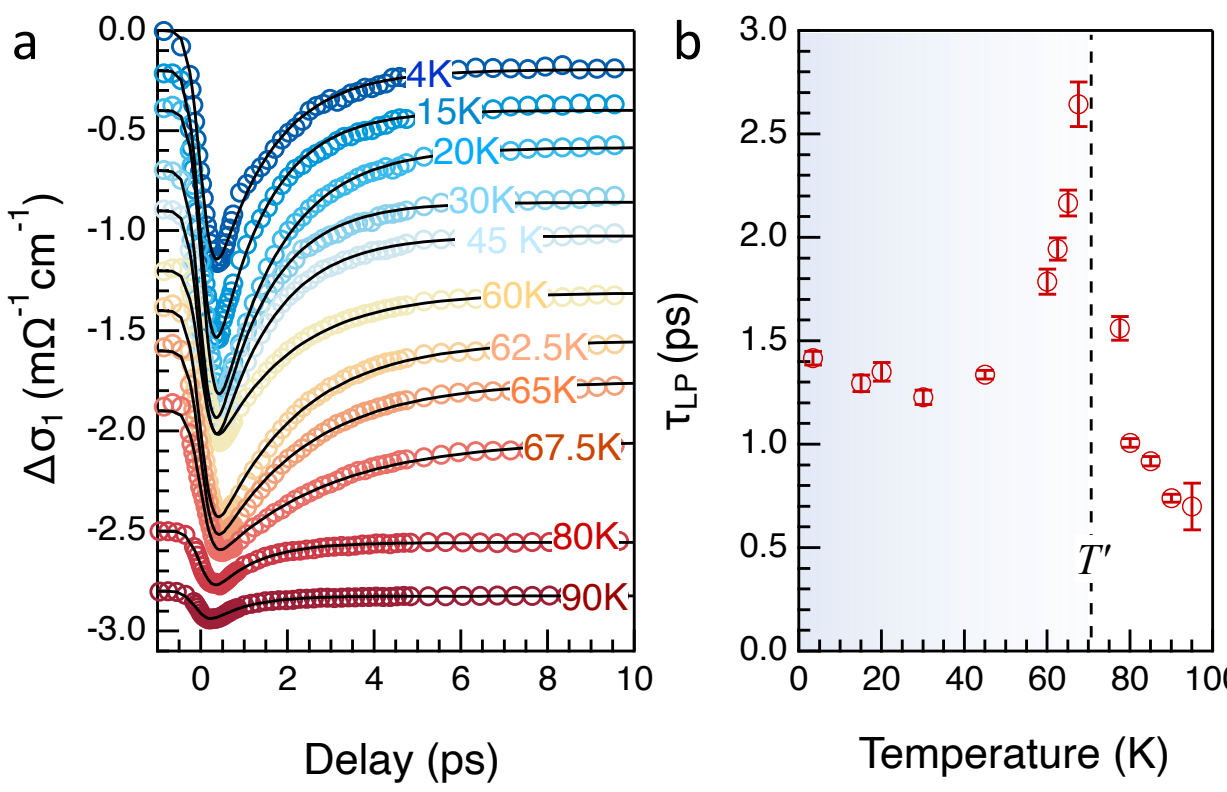

**Figure 4. Temperature dependent recovery kinetics of the non-superconducting component.** (a) Representative $\Delta\sigma_1$ kinetics in the low-$I_p$ regime ($I_p$ = 9.1 $\mu J \cdot cm^{-2}$) measured at indicated temperatures. (b) The plot of characteristic time ($\tau_{LP}$) of the $\Delta\sigma_1$ decay versus temperature. The dash line indicates the divergence of $\tau_{LP}$.

To examine the binding energy of the aforementioned local pairs, the local pairing dynamics represented by the $\Delta\sigma_1$ kinetics was measured at various temperatures below $T_c$. As shown in Fig. 4a, the decay rate of $\Delta\sigma_1$ kinetics first decelerates and then accelerates with increasing temperature. To explicitly show the temperature dependence, the $\Delta\sigma_1$ kinetics are fit by a biexponential function, and the characteristic time ($\tau_{LP}$) corresponding to the major decay component is plotted against temperature in Fig. 4b. $\tau_{LP}$ exhibits a clear divergence near 70 K, and this critical temperature is denoted as $T'$. As demonstrated afore, these $\Delta\sigma_1$ kinetics exclusively measure the local pairing, whereas this divergent behavior near $T'$ is reminiscent of the divergence of the temperature-dependent superfluid recovery time near $T_c$.(*26, 39, 45, 58-60*) Furthermore, the temperature-dependent $\Delta$ is generally manifested as the characteristic divergence of the superfluid recovery time near $T_c$.(*26, 36, 39, 45, 58-60*) From a phenomenological perspective, the divergence of $\tau_{LP}$ for the local pairing near $T'$ may also suggest a temperature-dependent collective energy gap ($\Delta'$), and the binding energy of the local pairs is equal to $2\Delta'$. As $\Delta'$ closes at temperature above $T'$, the individual quasiparticles rather than local pairs should dominate the normal state in the interval of $T'$< $T$ < $T_c$. Indeed, the $\Delta\sigma_1$ decay rate in this temperature interval is independent of the pump fluence, and their decay pattern resembles the intraband thermalization (Fig. S6). Since the normal state tends to comprise more local pairs with decreasing temperature below $T'$, the $\Delta\sigma_1$ decay rate below $T'$ becomes more sensitive to $I_p$, and a typical biparticle decay pattern emerges when temperature is significantly lower than $T'$.

There are several signatures suggesting Δ′ < Δ below $T'$. First, $T'$ is substantially lower than $T_c$. Second, considering that the pairing time is inversely proportional to the energy gap,(*24, 39, 58*) when both local and superconducting pairs are split, the former recovers much slower than the latter (see Fig. 3c and 3d). Third, the local pairs seem more vulnerable than the superconducting pairs because at low $I_p$ the former are broken while the latter is intact. This selective excitation of the local pairs under the low-$I_p$ condition probably reflects the remarkably different binding energy between the condensed Cooper pairs and local pairs. In the low-$I_p$ scenario, the nonthermal HFPs generated from hot carrier relaxation quickly decays into lower frequency phonons (LFPs) due to the anharmonicity of lattice vibrations (Fig. S7). If this process is faster than HFP mediated Cooper pair breaking, then we will not observe the depletion of the superfluid component in the THz conductivity. In addition to HFPs decay, the LFPs may also be generated from hot carrier relaxation. As the occupation number of LFPs should be much greater than the HFPs, those LFPs with energy in the interval between 2Δ and 2Δ′ can then selectively break the local pairs. In the high-$I_p$ scenario, nonthermal HFPs could be built up since the HFP emission from the hot carriers surpasses its anharmonic decay (which likely occurs in the presence of high-density hot carriers)(*61*), and then the breaking of condensed Cooper pairs is observed.

We note that the local pairs below $T'$ should be distinct from the preformed Cooper pairs that are generally discovered outside the superconducting dome (e.g., in the pseudogap regime) in cuprate superconductors. In striking contrast to the temperature-dependent local pairing observed here, the time constant for QPs* relaxation into preformed pairs is usually independent of temperature.(*24, 39*) Although it has been reported that Δ may coexist with a pseudogap below $T_c$, the characteristic time for quasiparticle pairing across the temperature-independent pseudogap is also independent of the temperature.(*39*) Nevertheless, these local pairs share similarities with short-range PDW state observed in cuprates.(*10, 15, 17, 62*) First, the PDW state was also found coexisting with the superconducting Cooper pairs at temperatures well below $T_c$.(*9, 13, 15, 63*) Second, the PDW state also exhibits coherence on a length scale that is long enough to form an energy gap but is insufficiently long to sustain superfluidity.(*15, 17, 62, 64*)

In summary, we investigated the dynamics of the non-superconducting state in optimally doped YBCO superconducting films were interrogated using transient THz spectroscopy and unveiled evidence for local pairs coexisting with superconducting pairs. As the superconducting gap arising from the long-range coherent pairing is generally manifested as a characteristic divergence of superfluid recovery time near $T_c$, in this work, opening of an energy gap induced by local pairing with short-range coherence is invoked to interpret the temperature-dependent local pairing dynamics and the clear divergence near 70 K (remarkably lower than $T_c$). To explore the phase diagram for local pairs below $T_c$, extending the investigation of the temperature-dependent local pairing dynamics to cuprates with various hole doping densities is motivated as further work.

**Acknowledgements.** Y.Y. acknowledges the National Natural Science Foundation of China under Grant Nos. 22473093 and 22175145, National Key Research and Development Program of China (2022YFB3803304), Fundamental Research Funds for the Central Universities under Grant number 20720220011 and 20720240150.